\documentclass[prl,twocolumn,superscriptaddress]{revtex4-2}
\usepackage{graphicx}
\usepackage{amsfonts}
\usepackage{amsmath}
\usepackage{amssymb}
\usepackage{bm}
\usepackage{slashed}
\usepackage{dsfont}
\usepackage{lineno}
\usepackage{color}
\usepackage{soul}

\begin{document}
	\title{Effective impurity behavior emergent from non-Hermitian proximity effect}
	\author{Deguang Wu}
	\affiliation{National Laboratory of Solid State Microstructures and Department of Physics, Nanjing University, Nanjing 210093, China}
	\author{Jiasong Chen}
	\affiliation{National Laboratory of Solid State Microstructures and Department of Physics, Nanjing University, Nanjing 210093, China}
	\author {Wei Su}
	\affiliation{College of Physics and Electronic Engineering, Center for Computational Sciences, Sichuan Normal University, Chengdu 610068, China}
	\affiliation{Beijing Computational Science Research Center, Beijing 100084, China}
	\author{Rui Wang}
	\email{rwang89@nju.edu.cn}
	\affiliation{National Laboratory of Solid State Microstructures and Department of Physics, Nanjing University, Nanjing 210093, China}
	\affiliation{Collaborative Innovation Center for Advanced Microstructures, Nanjing 210093, China}
	\author{Baigeng Wang}
	\affiliation{National Laboratory of Solid State Microstructures and Department of Physics, Nanjing University, Nanjing 210093, China}
	\affiliation{Collaborative Innovation Center for Advanced Microstructures, Nanjing 210093, China}
	\author{D. Y. Xing }
	\affiliation{National Laboratory of Solid State Microstructures and Department of Physics, Nanjing University, Nanjing 210093, China}
	\affiliation{Collaborative Innovation Center for Advanced Microstructures, Nanjing 210093, China}

	\maketitle
	{	\noindent
		\textbf{Abstract}
		
		\noindent
		Non-Hermitian boundaries commonly take place in many open quantum systems locally coupled to a surrounding environment. Here, we reveal a type of non-Hermitian effect induced by non-Hermitian boundaries, the non-Hermitian proximity effect (NHPE), which describes the penetration of non-Hermiticity from the boundary into the bulk. For gapped quantum systems, the NHPE generates in-gap states with imaginary eigenenergies, termed ``imaginary in-gap states". The imaginary in-gap states are localized at the system boundary and decay into the bulk, analogous to the behaviors of the conventional impurity states. However, in contrast to impurity states, the imaginary in-gap states exhibit distinct dynamical behaviors under time-evolution. Moreover, they are physically manifested as corner modes under open boundaries, as a combined result of the non-Hermitian skin effect (NHSE) and NHPE.  These results not only uncover implicit similarities between quantum systems with non-Hermitian boundaries and impurity physics, but also point to intriguing non-Hermitian phenomena broadly relevant to open quantum systems.}
	
	\vspace{2ex}
	\noindent
	\textbf{Introduction}
	
	\noindent
	Quantum systems that interact with an environment are ubiquitous in physics. The study of the interaction between such systems and baths is of fundamental interest as it leads to various novel quantum effects \cite{breuer2002theory,landi2022nonequilibrium} and applications  \cite{carmichael2009open}. In open quantum systems, although the system and the bath as a whole is Hermitian, the dynamics of the partial system alone can be described by an effective non-Hermitian model \cite{rotter2009non}. This has recently aroused enormous interest in uniform non-Hermitian quantum systems, where exotic physics have been found, including novel topological phases \cite{bender1998real,esaki2011edge,lee2016anomalous,yao2018non,kunst2018biorthogonal,
		kawabata2018anomalous,gong2018topological,lee2019topological,wang2019non,kawabata2019symmetry,luo2019higher,zhang2020correspondence,roccati2023hermitian}, skin effects \cite{yao2018edge,lee2019anatomy,lee2019hybrid,longhi2019probing,okuma2020topological,yi2020non,kawabata2020higher,okuma2021non,sun2021geometric,li2022gain,zhang2022review,roccati2021non}, enriched classifications \cite{shen2018topological,zhou2018observation,yin2018geometrical,yokomizo2019non,song2019non,deng2019non,kawabata2020non,yang2020non,kawabata2021topological,xue2021simple,wu2022connections,rui2022non}, and  unusual critical phenomena \cite{yoshida2018non,song2019non1,zhou2020renormalization}.
	
	It is however important to note that non-Hermiticity in realistic open quantum systems can be non-uniform, and in many cases, is only present at the system boundary. We consider systems interacting with their surrounding environments via short-range couplings across the boundaries. The general Hamiltonian reads as $H=H_S+H_E+V_{SE}$, where $H_S$ describes the closed quantum system, $H_E$ captures the environment consisting of a continuum of scattering wavefunctions, and $V_{SE}$ describes the coupling between them, which is short-ranged and restricted to the region around the boundary. The renormalization of quantum system from the environment via $V_{SE}$ is non-negligible \cite{rotter2009non}, which formally leads to the effective Hamiltonian as
	\begin{equation}\label{eq0}
		H_{eff}=H_S+\sum_{E} V_{SE}\frac{1}{\omega^+-H_E}V^{\dagger}_{SE},
	\end{equation}
	where $\omega^+=\omega+i0^+$ and the second term includes the sum over all the scattering channels in the environment. This term arising from system-bath coupling is clearly non-Hermitian. Importantly, since $V_{SE}$ is restricted in real space around the boundary, the non-Hermitian terms take place only at the boundary. The non-Hermitian boundaries implicit in Eq. \eqref{eq0} are common in open quantum systems. However, their effects have not received enough attention in recent studies. Could new phenomena emerge from the non-Hermitian boundaries, and how the non-Hermitian boundaries would affect the quantum systems and their dynamics? These key questions are yet to be addressed.
	
	In this work, we reveal a effect induced by non-Hermitian boundary, i.e., non-Hermitian proximity effect (NHPE). We show that the non-Hermiticity of the boundary can penetrate into the nearby quantum systems with a finite penetration depth, akin to the proximity effect of superconductors \cite{meissner1960superconductivity,liu2011abelian,xu2014artificial,zhu2016proximity}. For gapped quantum systems, the NHPE induces ``in-gap states" with imaginary eigenenergies, i.e., imaginary in-gap states. The imaginary in-gap states display peaks of local density of states (LDOS) clearly distinct from that of the bulk states. In addition, the imaginary in-gap states decay into the bulk in a way phenomenologically similar to that of conventional impurity problems \cite{yu1965bound,shiba1968classical,rusinov1969superconductivity,wang1996impurity,hyman1997impurity,balatsky2006impurity}, where the localized in-gap states decay into the bath \cite{balatsky2006impurity,yin2015observation,zheng2015interplay,sun2016majorana}. Despite the similarity, the imaginary in-gap states also exhibit several distinct properties. First, the imaginary in-gap states are generally manifested as corner modes under open boundary conditions (OBCs), due to the combined effect of the NHPE and the non-Hermitian skin effect (NHSE). Second, the NHPE leads to unusual dynamical behaviors of the quantum system. In particular, in the cases where the imaginary part of the imaginary in-gap state eigenenergy is positive, the probability distribution of imaginary in-gap state keeps increasing due to the gain from environment. As a result, under time-evolution, all wave-packets will be evolved into the imaginary in-gap states displayed as corner modes under OBCs.
	Our work reveals that impurity-like behaviors with unusual dynamics can emerge from quantum models with non-Hermitian boundaries. This points to a new direction that interwinds impurity physics and non-Hermitian effects, which could commonly take place in open quantum systems.
	
	\vspace{2ex}
	\noindent
	\textbf{Results}
	
	\noindent
	{\bf Imaginary in-gap states.}
	\begin{figure}[t]
		\includegraphics[width=\linewidth]{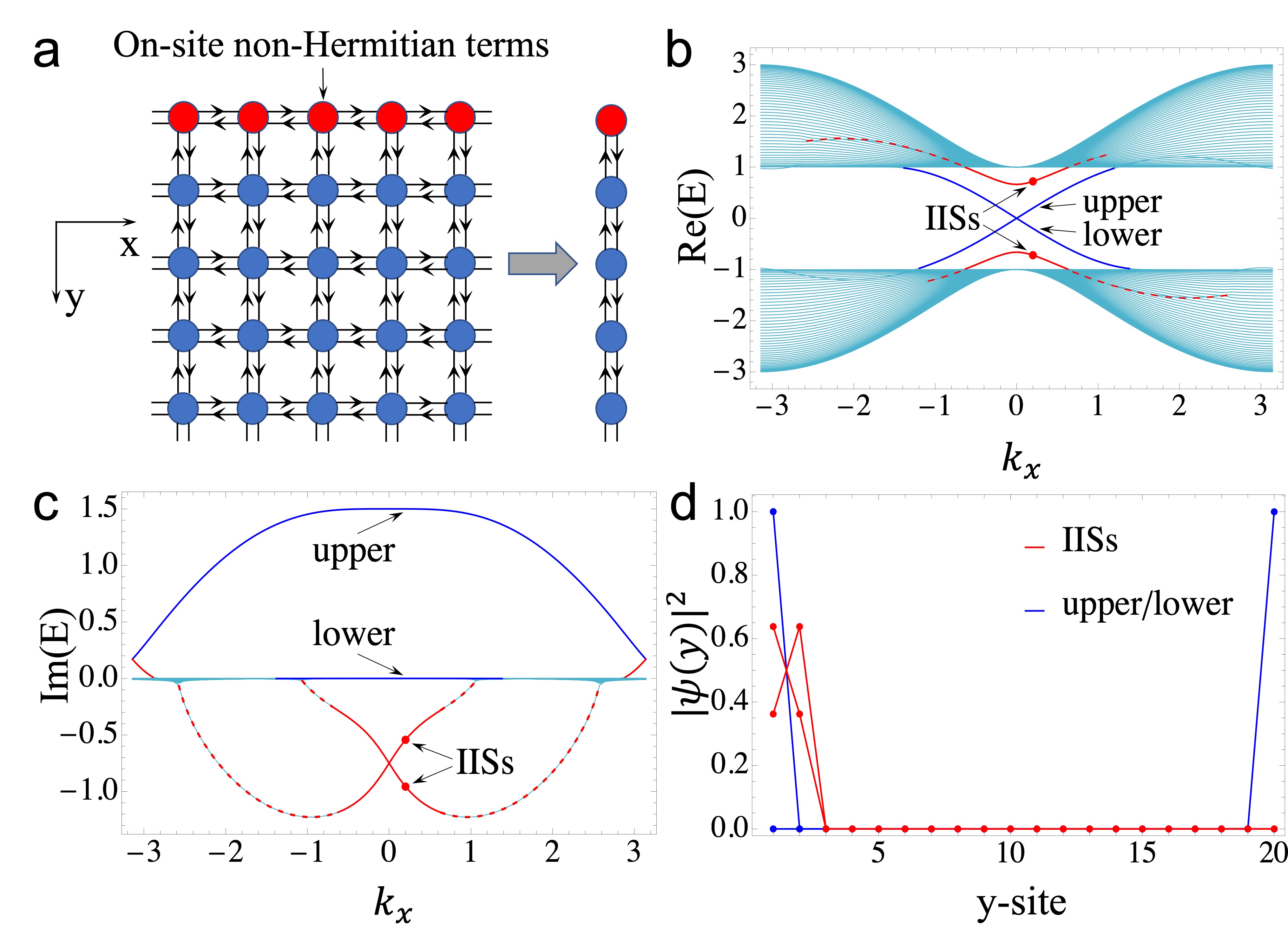}
		\caption{\textbf{Imaginary in-gap states (IISs) induced by non-Hermitian boundary.} \textbf{a} The Qi-Wu-Zhang (QWZ) model with on-site non-Hermitian boundary terms $i\gamma\sigma_x$ (marked by red circles) on the top layer of the system and on-site mass terms $m\sigma_z$ (marked by blue circles). Under the cylinder geometry (the size of the system in the $y$-direction is denoted by $L_y$), for any fixed $k_x$, the model is described by a 1D vertical chain with a non-Hermitian term on its first site. \textbf{b-c} The real and imaginary parts of the spectrum for the model $H=H_0+H_{\gamma}$ under the cylinder geometry, where $H_0$ is given by Eq. (\ref{eq1}), and $H_\gamma$ is given by Eq. (\ref{eq2}). The imaginary in-gap states expand into the bulk states as ``tail" states marked by the red dashed curves. The parameters used are $m=1.0$, $\gamma=1.5$ and $L_y=20$. \textbf{d} The probability distribution of the imaginary in-gap states and the topological chiral edge states  with $k_x=0.2$, which are marked by the arrows in (\textbf{b}) and (\textbf{c}).}\label{Fig1}
	\end{figure}
	We exemplify our study by starting with the Qi-Wu-Zhang (QWZ) model \cite{qi2006topological} describing 2D Chern insulators on a square lattice (Fig. \ref{Fig1}a), i.e.,
	\begin{equation}\label{eq1}
		\begin{split}
			H_0&=\sum_{x,y}\Bigl\{ \Bigl[c^{\dagger}_{x+1,y}\frac{(i\sigma_x-\sigma_z)}{2}c_{x,y} +c^{\dagger}_{x,y+1} \\ &\frac{(i\sigma_y-\sigma_z)}{2}c_{x,y}\Bigr]+
			c^{\dagger}_{x,y}m\sigma_zc_{x,y}\Bigr\}+h.c.,
		\end{split}
	\end{equation}
	where $c(x,y)=[c_{a,x,y},c_{b,x,y}]^{\mathrm{T}}$ is the spinor with $a$, $b$ sublattice.  We further consider the non-Hermitian terms along the boundary $y=1$:
	\begin{equation}\label{eq2}
		H_{\gamma}=i\sum_xc^{\dagger}_{x,1}\gamma\sigma_xc_{x,1}.
	\end{equation}
	Eq. \eqref{eq2} describes a non-Hermitian boundary as indicated by the red sites in Fig. \ref{Fig1}a, which can be generated by coupling $H_0$ to an environment at $y=1$.
	
	We first assume the cylinder geometry, i.e., the periodic boundary condition along $x$ and OBC along $y$ with $y\in[1,L_y]$. The real and imaginary part of energy spectrum are numerically obtained and shown in Figs. \ref{Fig1}b and c, respectively. Two chiral edge states of the Chern insulator are observed in the real spectrum, connecting the conduction and valence band. These two edge states are respectively localized at the upper ($y=1$) and lower ($y=L_y$) boundary, as shown by Fig. \ref{Fig1}d. It is also shown in Fig. \ref{Fig1}c that the upper chiral mode displays nonzero imaginary spectrum while the lower mode remains real. This is expected since the non-Hermitian term $\gamma$ is only applied on the upper boundary.
	
	Despite the chiral edge states, we also observe two channels of in-gap modes for $-0.6\lesssim k_x \lesssim0.6$, as marked by the red solid curves in Fig. \ref{Fig1}b. These in-gap modes do not connect the conduction and valence band, suggesting that they have a non-topological physical origin. Both channels describe edge states localized at the upper boundary, as evidenced by Fig. \ref{Fig1}d. Moreover, as shown by Fig. \ref{Fig1}c, each mode with fixed $k_x$ (marked by a red dot) in the two edge states exhibits nonzero imaginary eigenenergy, thus is termed as the imaginary in-gap state. We note that the dispersion of imaginary in-gap states has finite extensions into the bulk states with displaying ``tail" states, as marked by the red dashed curves in Fig. \ref{Fig1}b.

\begin{figure}
		\includegraphics[width=\linewidth]{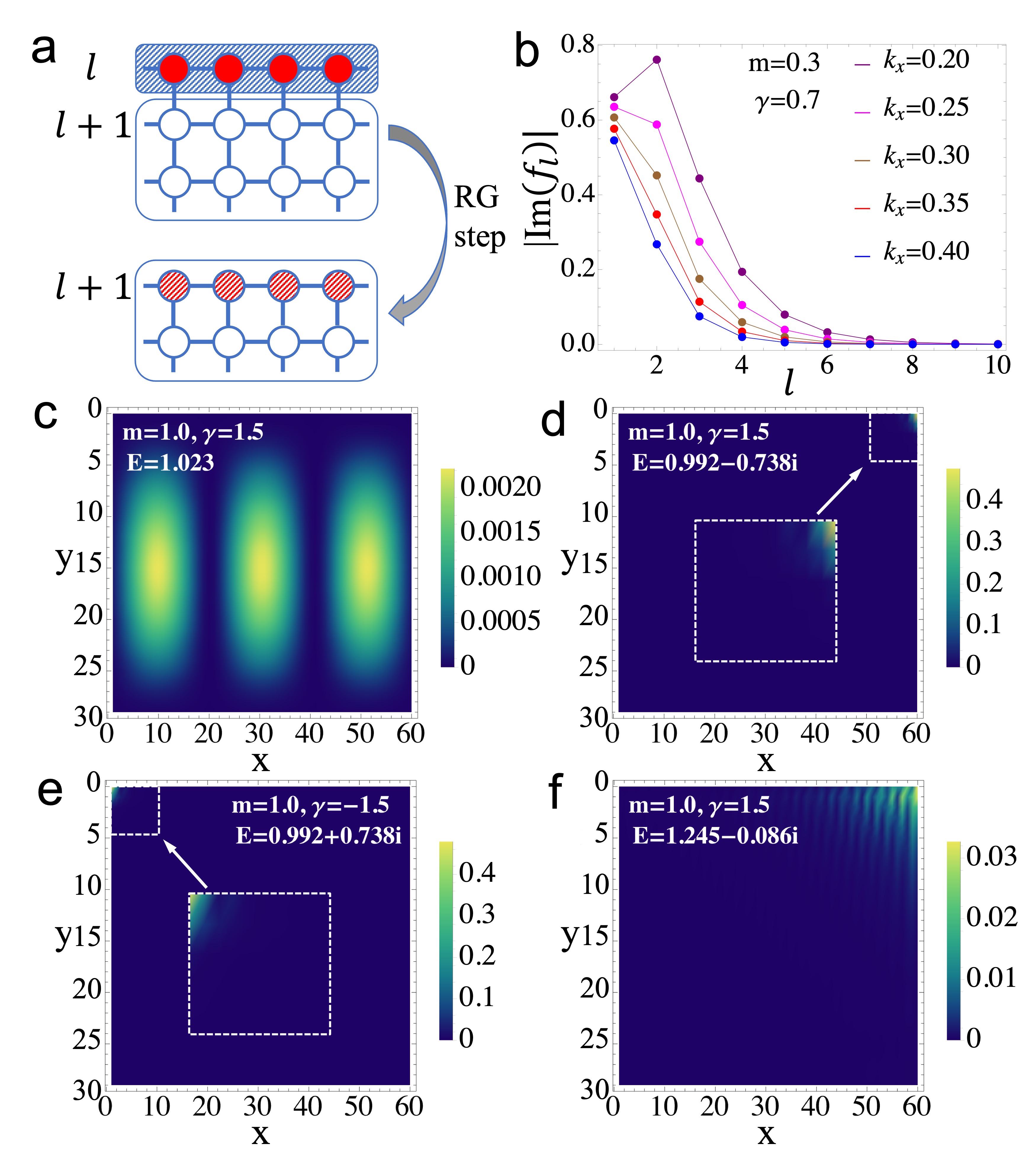}
		\caption{\textbf{Non-Hermitian proximity effect and the corner modes under open boundary conditions (OBCs).} \textbf{a} Schematic plot of the real-space renormalization group (RG) procedure. The renormalized ($l$+1)-th layer can be obtained after integrating out the $l$-th layer. \textbf{b} The imaginary part of the complex self-energy $f_l$ as a function of the RG step $l$ for different $k_x$. $\mathrm{Im}[f_l]$ decays with $l$ and eventually saturates to the fixed point $\mathrm{Im}[f_l]=0$ for large $l$. The model we use is $H=H_0+H_{\gamma}$ where on-site non-Hermitian boundary terms $i\gamma\sigma_x$ are on the top layer of the system. \textbf{c-f} The probability distribution of eigenstates under OBCs with size $L_x\times L_y=60\times30$, the colour bars on the right side of the panels indicate its value under normalization. (\textbf{c}) shows the case for the bulk states, (\textbf{d})-(\textbf{e}) are results for the imaginary in-gap states with opposite values of $\gamma$, and (\textbf{f}) shows the result for the ``tail state" deep in the bulk.}\label{Fig2}
	\end{figure}

	\vspace{2ex}
	\noindent
	{\bf Non-Hermitian proximity effect.} To reveal the origin of imaginary in-gap states, we first perform a real-space renormalization group (RG) analysis (see Supplementary Note 1). We decompose the total system $H=H_0+H_{\gamma}$ into a series of coupled 1D horizental layers, i.e., $H=\sum_{k_x}[\sum^{L_y}_{l=1}H_{l,k_x}+\sum^{L_y-1}_{l=1}H_{l,l+1,k_x}]$,
	where $H_{l,k_x}$ is the Hamiltonian of the $l$-th layer, and $H_{l,l+1,k_x}$ depicts the coupling between the $l$-th and $(l+1)$-th layer. The non-Hermitian boundary, i.e., Eq. \eqref{eq2}, is included in $H_{1,k_x}$, which reads as, $H_{1,k_x}=(\sin k_x+i\gamma)\sigma_x+(m-\cos k_x)\sigma_z$. Then, we integrate out the $l=1$ layer and obtain a renormalized Hamiltonian for the rest of the system, whose top layer ($l=2$) now becomes non-Hermitian.  Then, as indicated by Fig. \ref{Fig2}a, by iteratively repeating the above procedure, the effective Hamiltonian for the $(l+1)$-th layer can be obtained after the $l$-th RG step as,
	\begin{equation}\label{eq3}
		H^{eff}_{l+1,k_x}=\sin k_x\sigma_x+(m-\cos k_x)\sigma_z+f_l(\sigma_0+\sigma_x),
	\end{equation}
	where $f_l$ is the complex self-energy with $k_x$-dependence. In Fig. \ref{Fig2}b, we plot the calculated $\mathrm{Im}[f_l]$, as a function of the RG step $l$. As shown, $\mathrm{Im}[f_l]$ decays with $l$ for all $k_x$, and saturates to the fixed point $\mathrm{Im}[f_l]=0$ for large $l$. This shows that the non-Hermitian terms on the boundary can effectively penetrate into the bulk with a penetration depth. Due to its analogy with the proximity effect of superconductivity where Cooper pairs leak into the adjacent system, we term it the NHPE.
	
	To better visualize the NHPE, we adopt OBCs along both $x$ and $y$ directions, and calculate the spatial distribution of the eigenstates of $H$. The typical spatial distribution of the bulk states is shown by Fig. \ref{Fig2}c. It is clear that the bulk states remain as Bloch states and are barely affected by the non-Hermitian effects. This is expected from Fig. \ref{Fig2}b, since the non-Hermitian terms are negligible for bulk states far away from the upper boundary. Then, we focus on the imaginary in-gap states and pick up their corresponding eigenstates under OBCs. The typical spatial distribution of imaginary in-gap states are shown in Figs. \ref{Fig2}d and e. Due to the NHSE, which is significant near $y=1$, the imaginary in-gap states are found to localize either at the right or left boundary, depending on the sign of $\gamma$ (Figs. \ref{Fig2}d and e respectively). Moreover, they are also localized at the upper boundary due to the NHPE, with a short localization length along $y$ up to a few layers. Therefore, the combination of NHSE and NHPE generally drives the imaginary in-gap states into corner modes under OBCs. In addition, we also plot in Fig. \ref{Fig2}f the distribution of the ``tail states" (the dashed curve in Fig. \ref{Fig1}b) deep in the bulk. Compared to the imaginary in-gap states, these states are found to exhibit longer localization length along $y$ because of their smaller imaginary part of eigenenergies.
	
	\vspace{2ex}
	\noindent
	{\bf Analogy between imaginary in-gap states and impurity states.}
	So far, we have shown that the imaginary in-gap states manifest themselves as edge modes and corner modes under the cylinder geometry and OBCs, respectively. For both cases, the imaginary in-gap states are localized at the upper boundary and decay along $y$, as a result of the NHPE induced by the non-Hermitian boundary.
	
	To further reveal the physical nature of imaginary in-gap states, we now demonstrate their underlying similarity with the conventional impurity states in gapped systems. Under the cylinder geometry, $k_x$ is a good quantum number. The total system can then be written as $H=\sum_{k_x}H_{k_x}$, and
	\begin{equation}\label{eq4}
		\begin{split}
			H_{k_x}&=\sum_y [c^{\dagger}_{k_x,y+1}T_yc_{k_x,y}+h.c.]\\
			&+\sum_y \mathcal{\epsilon}_{k_x}c^{\dagger}_{k_x,y} c_{k_x,y}+ic^{\dagger}_{k_x,1}\gamma\sigma_xc_{k_x,1},
		\end{split}	
	\end{equation}
	where $c_{k_x,y}=[c_{a,k_x,y},c_{b,k_x,y}]^{\mathrm{T}}$, $T_y=(i\sigma_y-\sigma_z)/2$ and $\mathcal{\epsilon}_{k_x}=[\sin{k_x}\sigma_x+(m-\cos{k_x})\sigma_z]$.  For fixed $k_x$, $H_{k_x}$ describes a 1D vertical chain model with a non-Hermitian term on its first site, as indicated by Fig. \ref{Fig1}a.  It is clear that the localization and decay behaviors of an imaginary in-gap state is fully captured Eq. \eqref{eq4}.
	
	
	We now calculate the LDOS from Eq. \eqref{eq4}. The density of states (DOS) can be obtained from the imaginary part of the retarded Green's function GF, i.e., $\rho(\epsilon)=-\frac{1}{\pi}\mathrm{Im}\mathrm{Tr}[\sum_n\frac{|\psi^R_n\rangle\langle \psi^L_n|}{\epsilon+i0^+-E_{n}}]$, where $E_n$ is the $n$-th complex eigenvalues. Here, compared to the Hermitian cases, the support of DOS is expanded to the complex plane \cite{brouwer1997theory,mudry1998density}, i.e., $\epsilon=\epsilon_r+i\epsilon_i$. Besides, both the left and right eigenstates have been used to form the complete basis \cite{Brody_2014} that expands the GF.  Then, the LDOS at the $l$-th site of the chain can be derived as $\rho_l(\epsilon_r,\epsilon_i)=\frac{1}{N}\sum_n\delta(\epsilon_r-\mathrm{Re}[E_n])\delta(\epsilon_i-\mathrm{Im}[E_n])|\langle l|\psi^R_n\rangle\langle\psi^L_n|l\rangle|$.
	
\begin{figure}
		\includegraphics[width=\linewidth]{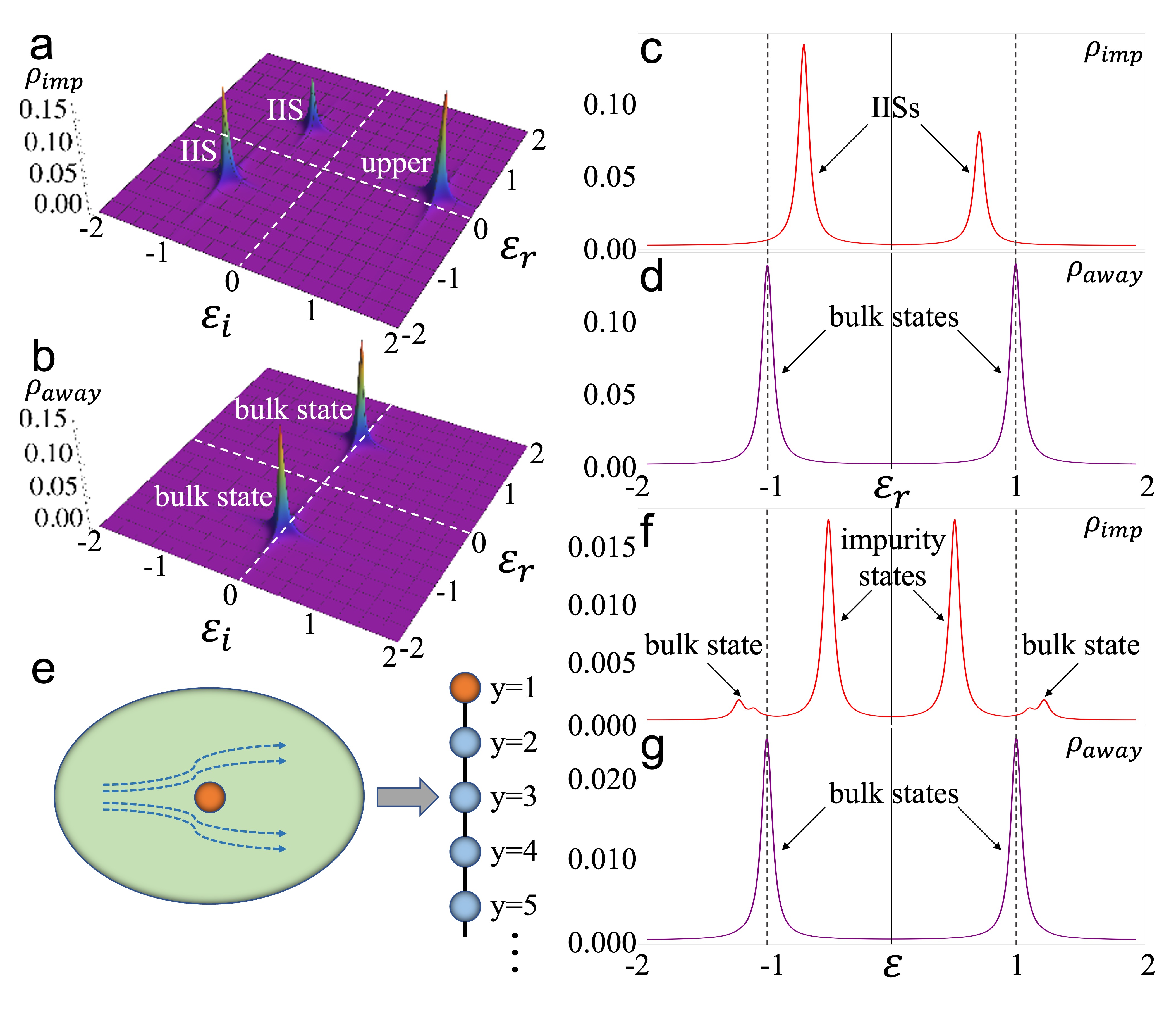}
		\caption{\textbf{Analogy between imaginary in-gap states and impurity states and their local density of states (LDOS).} \textbf{a-b} The LDOS of the non-Hermitian 1D vertical chain model (for fixed $k_x$) in Eq. (\ref{eq4}) with a non-Hermitian term $i\gamma\sigma_x$ on its first site  and on-site mass terms $m\sigma_z$. The parameters are $m=1.0$, $\gamma=1.5$, and $k_x=0.2$. In (\textbf{a}), there emerges two imaginary in-gap state (IIS) peaks together with that from the upper topological edge state at $y=1$. In (\textbf{b}), far away from the boundary ($y=7$), only the LDOS from bulk states is left which located on the real energy axis. \textbf{c} and \textbf{d} The projections of (\textbf{a}) and (\textbf{b}) onto the real axis. The LDOS from the topological edge state, which is not our focus, has been excluded for clarity. The gap energy is at $\omega=\pm1$, as marked by the dashed lines. \textbf{e} Schematic plot of an impurity embedded in the bath, which is mapped to an 1D Wilson open chain. \textbf{f-g} The calculated LDOS of the Wilson chain with the impurity level $\epsilon_f$ and the hybridization matrix element $V$. The parameters used are $m=1.0$, $\epsilon_f=0$, and $V=3.0$. In (\textbf{f}), two impurity in-gap states emerge near the impurity site, and two minor peaks also arise from the mixing with the bulk states. In (\textbf{g}), far away from the impurity, only the low-energy bulk states are left in the LDOS.}\label{Fig3}
	\end{figure}

	We show in Figs. \ref{Fig3}a and b the calculated LDOS, with focusing on the imaginary in-gap state region ($-0.6\lesssim k_x\lesssim0.6$). At the boundary site $y=1$ (Fig. \ref{Fig3}a) of the 1D vertical chain, except for the topological edge mode from the upper chiral edge state, we observe  two  significant LDOS peaks located at energies with $\epsilon_i<0$. These peaks come from the two imaginary in-gap states localized at $y=1$. Moving away from $y=1$, the imaginary in-gap state peaks quickly decay. Meanwhile, the LDOS from the bulk states emerges, which is located on the real energy axis. As shown by Fig. \ref{Fig3}b, for $y\gg1$, both the peaks from the imaginary in-gap states and the topological edge state disappear, leaving only the bulk states. The LDOS projected to the real energy axis is also shown in Figs. \ref{Fig3}c and d for clarity.
	
	To clearly show the similarity between imaginary in-gap states and impurity states, we consider a non-interacting pseudospin Anderson impurity model coupled to the QWZ model, i.e., $H_{imp}=H_0+H_{f}$, where the impurity is described by
	\begin{equation}\label{eqimpu}
		H_{f}=\sum_{\sigma}\epsilon_ff^{\dagger}_{\sigma}f_{\sigma}+V\sum_{\mathbf{k},\sigma}(f^{\dagger}_{\sigma}c_{\mathbf{k},\sigma}+h.c.).
	\end{equation}
	$f_{\sigma}$ is the annihilation operator of the impurity state with pseudospin (sublattice) $\sigma$, $\epsilon_f$ denotes the impurity energy level (independent of $\sigma$), and $V$ its hybridization with the bath electrons. Although the impurity considered in Eq. \eqref{eqimpu} is located in the bulk of the 2D Chern insulator as shown by Fig. \ref{Fig3}e, the low-energy impurity state is accurately determined by an 1D Wilson open chain following the numerical renormalization group mapping scheme (see Supplementary Note 4), as schematically shown by Fig. \ref{Fig3}e.
	\begin{figure}
		\includegraphics[width=\linewidth]{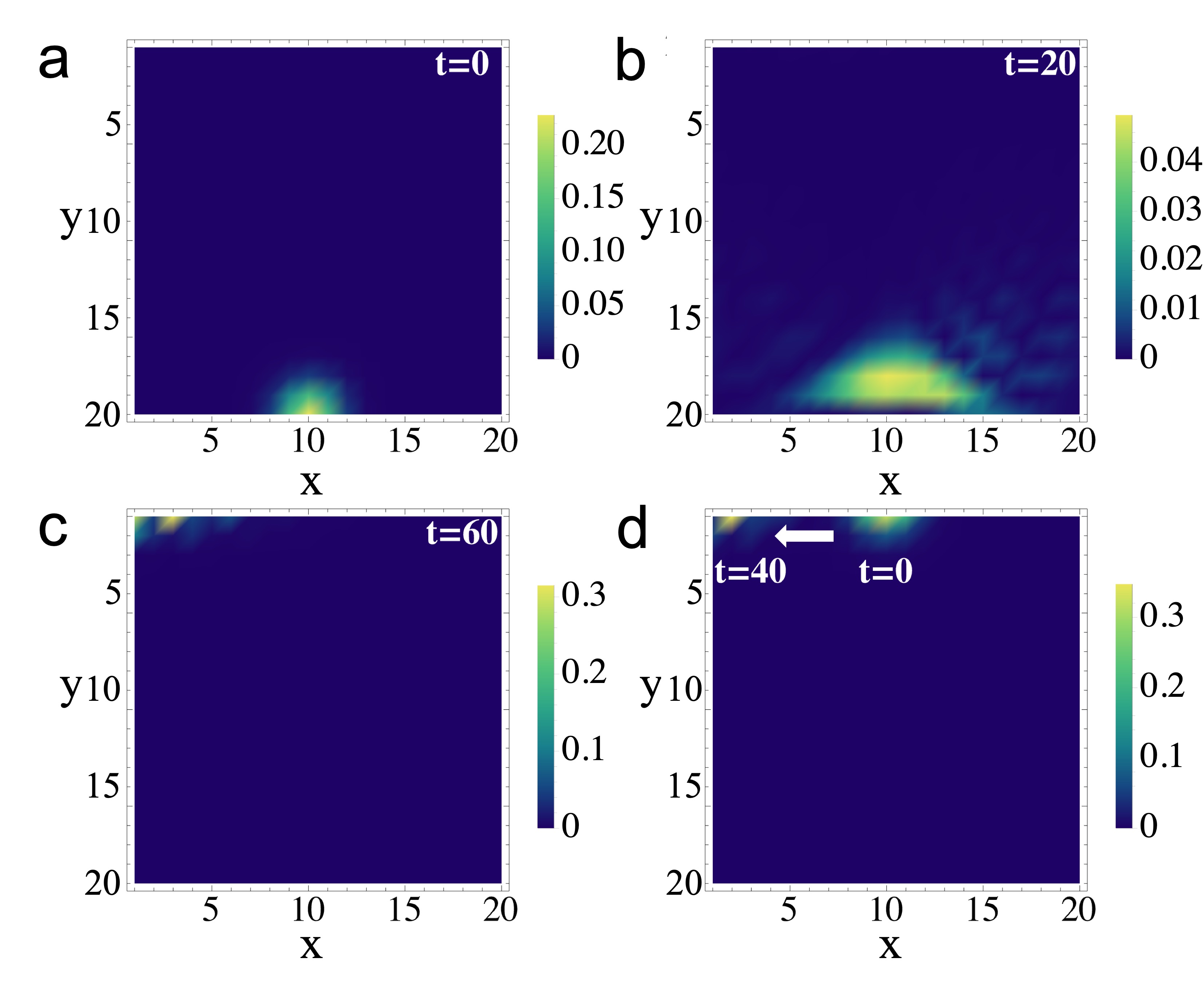}
		\caption{\textbf{Time-evolution of the wave-packet.} \textbf{a} The initial state is set to be a Gaussian wave-packet, placed at the bottom layer of a square lattice with size $L_x\times L_y=20\times20$. The model is described by $H=H_0+H_{\gamma}$, where $H_0$ is given by Eq. (\ref{eq1}) which contains on-site mass terms $m\sigma_z$, and $H_\gamma$ is given by Eq. (\ref{eq2}) with the on-site non-Hermitian boundary terms $i\gamma\sigma_x$ located on the top layer of the system. The wave-packet evolves according to the Schr\"{o}dinger equation. \textbf{b} The Gaussian wave-packet diffuses into the bulk. \textbf{c} The Gaussian wave-packet eventually evolves into the imaginary in-gap state localized at the upper-left corner. \textbf{d} The evolution of a wave-packet initially placed at the top layer. The colour bars on the right side of the panels indicate the value of the normalized probability distribution at different times. Common parameters: $m=1.0$, $\gamma=-1.5$.}
		\label{Fig4}
	\end{figure}
	The Wilson open chain shares a similar structure with the 1D vertical chain model in Fig. \ref{Fig1}a. Both of them describe a boundary site ($y=1$) coupled to bath degrees of freedom ($y>1$) via the nearest neighbor coupling.  It should be noted that the $y>1$ sites on the Wilson chain represent for the bath states in discretized energy windows. The site being closer to the impurity ($y=1$) describes the states in the higher energy window, i.e., in the shorter length scale measured from the impurity.
	
	We then calculate the LDOS on different sites of the Wilson chain, as shown in Figs. \ref{Fig3}f and g.  For $y$ being close to the impurity, two in-gap states emerge, due to the coupling of the impurity to the conduction and valence band electrons.  With tuning $y$ away from the impurity, the LDOS from the impurity states decreases and that from the bulk states increases. For large $y$, only the LDOS of the low-energy bulk states are left, which are located around the gap energy, as shown by Fig. \ref{Fig3}g. The LDOS shown in Figs. \ref{Fig3}f and g is qualitatively in analogy with the projected LDOS in Figs. \ref{Fig3}c and d, implying the similarity between the imaginary in-gap states and the impurity physics.
	
	\vspace{2ex}
	\noindent
	{\bf Distinct dynamical behaviors of imaginary in-gap states.}
	Despite the similarity, the imaginary in-gap states also exhibit some distinct features. First, since the LDOS peaks of imaginary in-gap states are located in the complex energy plane with $\epsilon_i\neq0$, they can persist even if their real parts of eigenenergies are immersed in the bulk continuum (see Supplementary Note 2). This explains the existence of the ``tail states" observed by Fig. \ref{Fig1}b.
	Second,  the imaginary in-gap states exhibit unusual dynamical behaviors absent in the Hermitian systems, as a result of their imaginary eigenenergies.  Under time-evolution, the amplitude of an imaginary in-gap state either grows or decays, depending on the sign of $\gamma$ (see Supplementary Note 3). This is a reflection of the non-Hermitian nature of the imaginary in-gap states, in contrast with the conventional impurity states.
	
	For a Gaussian wave-packet input at the bottom layer of system, the time-evolution of the wave-packet is calculated under OBCs and shown in Figs. \ref{Fig4}a-c  (for $\gamma<0$). The wave-packet first diffuses into the bulk (Fig. \ref{Fig4}b) and then gradually evolves into the imaginary in-gap state eigenstate localized at the top layer. Due to the NHSE, it finally turns into the corner state as shown in Fig. \ref{Fig4}c. We also start from a Gaussian wave-packet placed at any generic site at the top layer. As shown by Fig. \ref{Fig4}d, the wave-packet also evolves into the imaginary in-gap state located at the corner. Hence, we observe that, for $\gamma<0$ where the amplitude of imaginary in-gap states grows, a generic Gaussian wave-packet will always evolve into the cornered imaginary in-gap state (see Supplementary Note 3 for the $\gamma>0$ case). Such a dynamical feature is a clear distinction between imaginary in-gap states and conventional in-gap states.
	
	\vspace{2ex}
	\noindent
	\textbf{Discussion}
	
	\noindent
	This work reveals an impurity-like non-Hermitian phenomenon induced by non-Hermitian boundaries. Although the imaginary in-gap state constitute an edge mode, each individual imaginary in-gap state can be further explored to simulate impurity physics with novel non-Hermitian properties \cite{sukhachov2020non,li2021impurity}. For example, by properly introducing Hubbard interaction $U$ on the boundary, the corresponding 1D vertical chain model in Eq. \eqref{eq4} would bare similarities with the Wilson chain mapped from a finite $U$ Anderson model. Hence, Kondo-like behaviors could emerge but enriched by new features arising from the non-Hermiticity.
	
\begin{figure}
		\includegraphics[width=\linewidth]{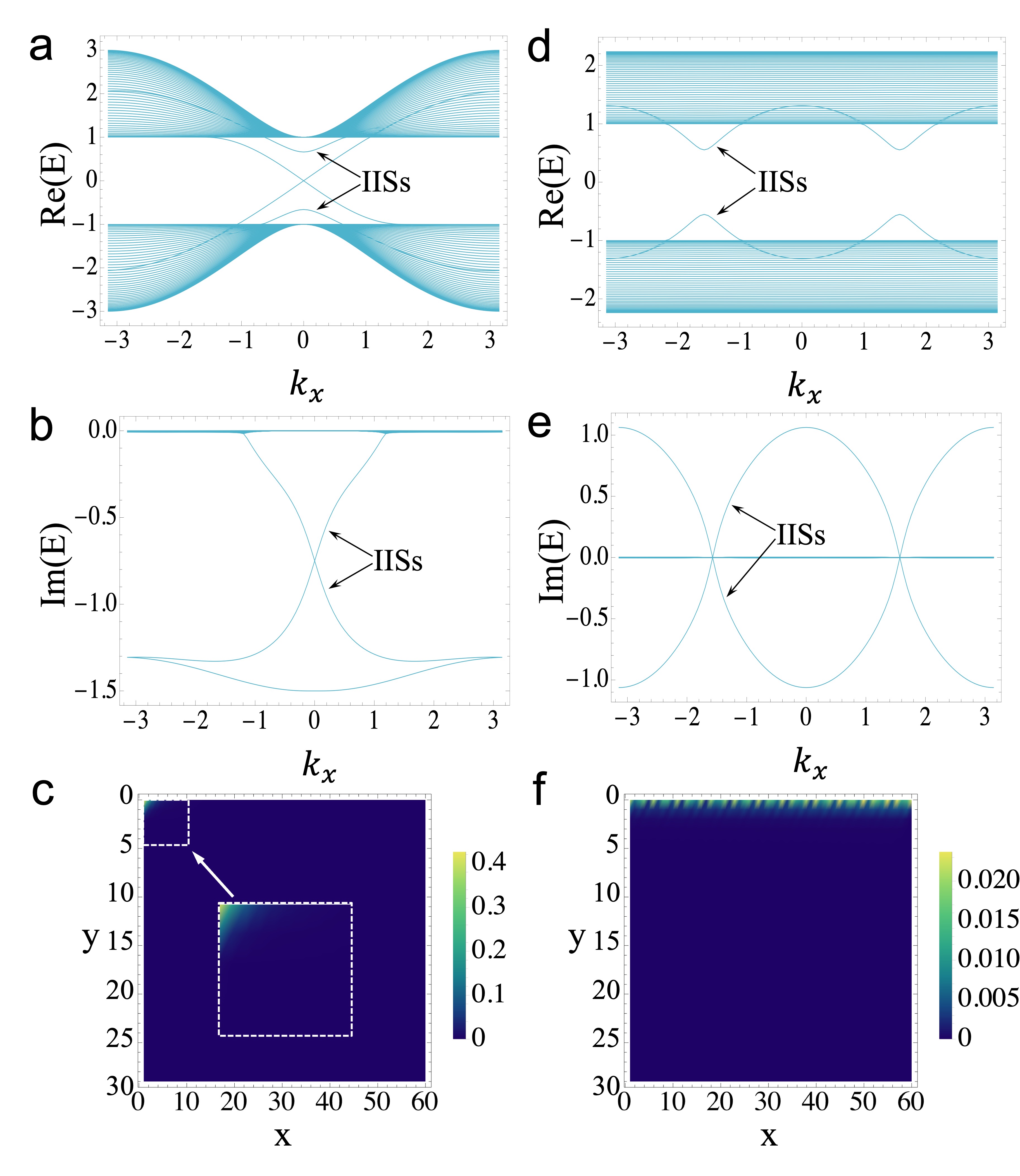}
		\caption{ \textbf{Generality of the non-Hermitian proximity effect (NHPE).} \textbf{a-b} The real and imaginary parts of the spectrum for the model $H=H_0+H_\gamma$, where $H_0$ is described by Eq. (\ref{eq1}) which contains on-site mass terms $m\sigma_z$, and $H_\gamma$ is described by Eq. (\ref{eqs29}) with the on-site non-Hermitian boundary terms $i\gamma\sigma_0$ located on the top layer of the system. The model obeys cylinder geometry with size in the $y$-direction being $L_y=50$.  The parameters used are $m=1.0$ and $\gamma=-1.5$. \textbf{c} The probability distribution of the imaginary in-gap states (IISs) under open boundary conditions (OBCs) with size $L_x\times L_y=60\times30$, where the imaginary in-gap state ($E=-0.451-1.222i$) is located at the corner due to the combined effects of the NHPE and the non-Hermitian skin effect (NHSE). The colour bar on the right side of the panel indicates the value of the probability distribution under normalization. \textbf{d-f} Analogous to (\textbf{a})-(\textbf{c}), with $H_0$ now being described by Eq. (\ref{eqs30}) and $H_\gamma$ now being described by Eq. (\ref{eq2}), where the parameters used are $m=1.0$ and $\gamma=1.5$. In (\textbf{f}), the imaginary in-gap state ($E=0.746+0.464i$) is an edge mode evenly distributed along the upper boundary, since only the NHPE exists in this model.}\label{Fig5}
	\end{figure}

	Although a non-Hermitian boundary with alternating gain and loss in Eq. \eqref{eq2} is studied as an example, as we show in Methods, the NHPE and the corresponding imaginary in-gap states remain intact for gain-only or loss-only boundaries as well. In addition, we also investigate a more general 2D insulator model with non-Hermitian boundaries. As shown in Methods, although the NHSE is absent in this model, the NHPE still persists, leading to imaginary in-gap states that are manifested by a localized edge mode shown in Fig. \ref{Fig5} (rather than the corner mode in Fig. \ref{Fig2}d). The model-independence indicates that the NHPE should be a more general non-Hermitian effect than NHSE.
	
	The theoretical model studied here could be realized in different experimental platforms. For example, the Chern insulator model $H$ can be realized by magnetically doped topological insulators \cite{yu2010quantized}. Moreover, the non-Hermitian boundary terms can be realized on the basis of reservoir engineering. The loss or gain on the boundary can be achieved by using a nonlocal coupling to auxiliary degrees of freedom which undergo local loss or gain \cite{gong2018topological}, as discussed in details in Methods.   Moreover,  the 2D topological insulating phases can also be realized by cold atoms in optical lattices or photons in coupled cavities \cite{reiter2012effective}. In particular, in arrays of coupled micro-ring cavities, the photon gain and loss for each cavity can be controlled independently \cite{zhao2018topological,mittal2019photonic}. These provide promising platforms to further investigate the predicted NHPE, the imaginary in-gap states, as well as their dynamical behaviors.
	
	\vspace{2ex}
	\noindent
	\textbf{Methods}
	
	\noindent
	\textbf{Real-space renormalization group analysis of non-Hermitian proximity effect.}
	For the open boundary perpendicular to the $y$-direction, the total system of a 2D lattice can be decomposed into a series of horizontal layers (labelled by $l$) with inter-layer coupling, and its partition function is then cast into $\mathcal{Z}=\int\prod_l\mathcal{D}\bar\psi_{l}
	\mathcal{D}\psi_{l}e^{-(S_l+S_{l,l+1})}$, where
	\begin{equation}\label{eqs1}
		S_l =-\sum_{i\omega_n}\sum_{k_x\in BZ}\bar\psi_{k_x,i\omega_n,l}
		[i\omega_n-H_{l}(k_x)]\psi_{k_x,i\omega_n,l},
	\end{equation}
	where $\psi_{k_x,i\omega_n,l}=[c_{a,k_x,iw_n},c_{b,k_x,iw_n}]^T$ is the Grassmann field. $H_l(k_x)$ is the Hamiltonian of the $l$-th horizental chain. The action for the inter-layer coupling can be written as
	\begin{equation}\label{eqs2}
		S_{l,l+1}=-\sum_{i\omega_n}\sum_{k_x\in BZ}(\bar\psi_{k_x,i\omega_n,l}T_y\psi_{k_x,i\omega_n,l+1}+h.c),
	\end{equation}
	where  $T_y$ is the matrix describing the inter-layer hopping, which is assumed to be $l$-independent. Since the action is bilinear in terms of Grassmann fields, we can integrate out the first layer and obtain an renormalized effective action for the second layer as
	\begin{equation}\label{eqs3}
		S_2=-\sum_{i\omega_n}\sum_{k_x\in BZ}\bar\psi_{k_x,i\omega_n,2}G^{-1}_{2}(i\omega_n,k_x)\psi_{k_x,i\omega_n,2},
	\end{equation}
	where $G^{-1}_{2}(\omega,k_x)=\omega^+-H_2(k_x)-T_y^{\dagger}(\omega^+-H_1(k_x))^{-1}T_y$
	is the renormalized retarded Green's function of the second layer with $\omega^+=\omega+i0^+$. To extract the low-energy effective Hamiltonian, the low-frequency approximation can be made in $G^{-1}_l(\omega,k_x)$,
	which well preserves the low-energy physics as long as the layer integrated out remains gapped (which is indeed the case for the QWZ model studied here, where each layer describes a 1D Su-Schrieffer-Heeger (SSH) model). Then, the effective Hamiltonian of the second layer is read off as
	\begin{equation}\label{eqs4}
		H_{2}^{eff}(k_x)=H_{2}(k_x)-T_y^{\dagger}\frac{1}{H_{1}(k_x)}T_y.
	\end{equation}
	By treating the renormalized second layer as the first layer on top of the remaining system, then the above procedure can be performed iteratively, leading to the effective action for the renormalized third, fourth... layer. After $l$ RG steps, an iterative relation between the effective Hamiltonian of the $l$-th and that of the ($l+1$)-th layer can be obtained as
	\begin{equation}\label{eqs5}
		H_{l+1}^{eff}(k_x)=H_l(k_x)-T_y^{\dagger}\frac{1}{H_l^{eff}(k_x)}T_y.
	\end{equation}
	
	Now we apply the real-space RG transformation to QWZ model with a non-Hermtian boundary, the first layer Hamiltonian with non-Hermitian terms is given by
	\begin{equation}\label{eqs8}
		H_1(k_x)=(\sin k_x+i\gamma)\sigma_x+(m-\cos k_x)\sigma_z,
	\end{equation}
	which is a non-Hermitian version of the SSH model and exhibits non-Hermitian skin effect \cite{yao2018edge}.
	We then derive the effective Hamiltonian for the renormalized $l$-th layer under the real-space RG transformation. For any $2\times2$ matrix A, it can be expanded into a linear combination, $A=a\sigma_0+b\sigma_x+c\sigma_y+d\sigma_z$, where $\sigma_{x,y,z}$ and $\sigma_0$ are the Pauli matrices and identity matrix, respectively. Then, we have
	\begin{equation}\label{eqs9}
		T_y^{\dagger}A^{-1}T_y=\frac{a+b}{2 \mathrm{Det}[A]}(\sigma_0+\sigma_x),
	\end{equation}
	with $ \mathrm{Det}[A]=a^2-(b^2+c^2+d^2)$. Using the Eqs. (\ref{eqs4}), (\ref{eqs8}) and (\ref{eqs9}), we have
	\begin{equation}\label{eqs10}
		H_2^{eff}(k_x)=\sin{k_x} \sigma_x+(m-\cos{k_x}) \sigma_z+f_1(\sigma_0+\sigma_x),
	\end{equation}
	where $f_1=(\sin{k_x}+i\gamma)/[2(2-2\cos{k_x}+2i\gamma\sin{k_x}-\gamma^2)]$ is the self-energy due to the renormalization of the first layer. Accordingly,  we can express the effective Hamiltonian of the $(l+1)$-th layer in the same form as Eq. (\ref{eqs10}), i.e., Eq. (\ref{eq3}). Substituting it into the Eqs. (\ref{eqs5}) and (\ref{eqs9}), an iterative relation between $f_{l+1}$ and $f_l$ can be obtained
	\begin{equation}\label{eqs12}
		f_{l+1}=\frac{\sin{k_x}+2f_l}{2(2\sin{k_x}f_l-2m\cos{k_x}+m^2+1)}.
	\end{equation}
	The fixed points, if exist, can be determined by requiring $f_{l+1}=f_l=f_c$ for $l\to\infty$. This leads to the following equation
	\begin{equation}\label{eqs13}
		4\sin{k_x}{f_c}^2-2m(2\cos{k_x}-m)f_c-\sin{k_x}=0.
	\end{equation}
	Since  $\varDelta=4m^2(2\cos{k_x}-m)^2+16\sin{k_x}^2\geqslant0$, the two roots $f_{c1}$ and $f_{c2}$ of Eq. (\ref{eqs13}) are given by
	\begin{equation}\label{eqs14}
		f_{c1/c2}=\frac{2m(2\cos{k_x}-m)\pm\sqrt{\varDelta}}{8\sin{k_x}},
	\end{equation}
	which implies that the self-energy $f_l$ are real at the fixed points. This explains why the imaginary part of $f_l$ eventually vanishes for large $l$, as shown in Fig. \ref{Fig2}b. In additional, we can prove that
	$f_l$ eventually flows to only one of the fixed points $f_{c1}$. More subtle details can be found in the Supplementary Note 1.
	
	\vspace{2ex}
	\noindent
	{\bf General existence of NHPE in gapped states with non-Hermitian boundaries.}
	{Although the non-Hermitian effects are derived from Eqs. \eqref{eq1} and (\ref{eq2}), which describe a Chern insulator with a boundary that has alternating gain and loss (on $a$ and $b$ sites), the NHPE found here is in fact  general. To demonstrate its generality, we first consider the non-Hermitian boundaries with only gain or only loss. Hence, we replace Eq. (\ref{eq2}) by
		\begin{equation}\label{eqs29}
			H_{\gamma}=i\sum_xc^{\dagger}_{x,1}\gamma\sigma_0c_{x,1},
		\end{equation}
		which describes the gain and loss on all the sites ($y=1$) for $\gamma>0$ and $\gamma<0$, respectively. We then calculate the real and imaginary part of energy spectrum of the total system
		$H=H_0+H_{\gamma}$ under cylinder geometry, which are respectively shown in Figs. \ref{Fig5}a and b. As we see from Figs. \ref{Fig5}a and b, two imaginary in-gap states still emerge when only loss is present, similar to the results for the boundaries with alternating gain and loss. By comparing Figs. \ref{Fig5}a and b with Figs. \ref{Fig1}b and c, we find that the only slight difference is that the imaginary part of the eigenenergies of the upper chiral edge mode now becomes negative. This does not affect the main conclusions on the imaginary in-gap states and the NHPE.}
	
	The general existence of NHPE can be understood in the RG sense. The RG analysis discussed above is general, and can be applied to gapped quantum systems with a generic non-Hermitian boundary. For instance, we consider the Eq. (\ref{eqs29}) being added to the QWZ model and perform the real-space RG analysis. Similar to the RG analysis for the case with alternating loss and gain, the bulk Hamiltonian can be decomposed into different layers coupled via hopping matrix $T_y$. The key fact here is that the iterative relation between the effective Hamiltonian of the $l$-th and that of the ($l+1$)-th layer, i.e., Eq. (\ref{eqs5}), does not change as long as $T_y$ remains the same. Thus, the effective Hamiltonian for the ($l+1$)-th layer can still be written as Eq. (\ref{eq3}) which leads to the same result as Eq. (\ref{eqs12}). This indicates that the different non-Hermitian boundary terms only leads to different initial value $f_1$, and the RG flow of $f_l$ remains qualitatively the same, independent of $f_1$. Consequently, the same fixed point with $\mathrm{Im}[f_l]=0$ will always be reached for large $l$. This means that NHPE is general and does not depend on the form of non-Hermitian boundary.
	
	Since the RG analysis shows that the non-Hermitian effect induced by the boundary would always decay and vanish for large distances from the boundary, the effects from different non-Hermitian boundaries would become independent from each other for systems larger than the decay length. Therefore, similar non-Hermitian phenomena is expected if non-Hermitian terms are considered on both the upper and lower boundaries.
	
	To further support the RG results, we also calculate the typical spatial distribution of imaginary in-gap state under OBCs. As shown in the Fig. \ref{Fig5}c, corner states still emerge for the gain-only or loss-only boundaries, as a result of the combined effect of the NHPE and NHSE, indicating the general existence of NHPE.
	
	We now show that the NHPE is not only independent of the specific forms of the boundary, but is also independent of the specific models of the insulating bulk. Instead of the QWZ model, we now consider a more general 2D two-band model with a bulk gap $m$ described by the following Hamiltonian:
	\begin{equation}\label{eqs30}
		H_0(k_x,k_y)=\sum_{i=x,y}(\cos{k_i}\sigma_x+\sin{k_i}\sigma_y)+m\sigma_z,
	\end{equation}
	along with the non-Hermitian boundary term $H_\gamma=i\sum_{x}{c^\dag_{x,1}\gamma\sigma_x}c_{x,1}$, i.e., the Eq. (\ref{eq2}). Then we calculate the energy spectrum under cylinder geometry. Figs. \ref{Fig5}d and e show the real and imaginary part of the energy spectrum, which clearly indicate the emergence of the imaginary in-gap states, similar to those for the QWZ model.
	
	As discussed above, the general existence of NHPE can be proved by real-space RG calculations. We therefore decompose the total system $H=H_0+H_\gamma$ (where $H_0$ is given by Eq. (\ref{eqs30}) above) into a series of coupled 1D horizontal layers. The Hamiltonian of the $l=1$ layer reads as $H_{1,k_x}=(\cos{k_x}+i\gamma)\sigma_x+\sin{k_x}\sigma_y+m\sigma_z$ and the inter-layer hopping matrix is $T_y=\sigma^+$. Then, we can derive the renormalized Hamiltonian for the ($l+1$)-th layer by iteratively integrating out the $l$-th layer, leading to:
	\begin{equation}\label{eqs31}
		H_{l+1,k_x}^{eff}=\cos{k_x}\sigma_x+\sin{k_x}\sigma_y+m\sigma_z+f_l(\sigma_0-\sigma_z),
	\end{equation}
	where complex self-energy $f_l$ encodes the non-Hermitian effect. Similarly, it can be proved that $f_l$ always saturates to the same fixed point with $\mathrm{Im}[f_l]=0$ for large $l$. This clearly suggests the boundary-induced non-Hermitian effect decays and finally vanishes deep in the bulk for large $l$. Since Eq. (\ref{eqs30}) is the minimal model of 2D insulators without any specific requirements, the NHPE found here is expected to be a general result for gapped quantum systems with a non-Hermitian boundary. This is also consistent with the emergence of imaginary in-gap states shown in Figs. \ref{Fig5}d and e.
	
	In particular, we mention that the NHSE is model-dependent and it does not occur in the above 2D insulator model. However, the NHPE remains intact. Correspondingly, the distribution of imaginary in-gap state calculated under open boundaries no longer shows up as a corner mode, but is manifested by an edge mode evenly distributed along the upper boundary (Fig. \ref{Fig5}f). This indicates that the NHPE could be a more general non-Hermitian effect than the NHSE.
	
	\vspace{2ex}
	\noindent
	{\bf Local density of states of the non-Hermitian vertical chain model.}
	In the Hermitian systems, the DOS can be calculated from the imaginary part of the retarded Green's function $\rho(\epsilon)=-\frac{1}{\pi}\text{Im Tr} [\sum_{n}\frac{ |\psi_n\rangle \langle\psi_n|}{\epsilon+i\eta-E_n}]$, where $\eta=0^+$, $H|\psi_n\rangle=E_n|\psi_n\rangle$, $n=1,2,\dots,N$. For the LDOS at the $l$-th site, we have
	\begin{equation}{\label{eqs23}}
		\begin{aligned}
			\rho_l(\epsilon)
			&=-\frac{1}{\pi}\text{Im}\displaystyle \left[\sum_{n}\frac{ \langle l|\psi_n\rangle \langle\psi_n|l\rangle}{\epsilon+i\eta-E_n}\right]\\
			&=\frac{1}{\pi}\sum_{n}\frac{ \eta\langle l|\psi_n\rangle \langle\psi_n|l\rangle}{(\epsilon-E_n)^2+\eta^2}\\
			&=\sum_{n}\delta(\epsilon-E_n)|\langle l|\psi_n\rangle|^2.
		\end{aligned}
	\end{equation}
	The last step in the above equation exploits the fact that the Dirac delta function is the limit of the Lorentzian function with $\eta$ going to zero, and the final result goes back to the definition of the DOS. However, Eq. (\ref{eqs23}) no longer holds for non-Hermitian Hamiltonians, since $E_n$ is now complex which makes it invalid. Thus,  the way to calculate the LDOS has to be generalized to capture the non-Hermitian systems \cite{brouwer1997theory,mudry1998density}. For a non-Hermitian Hamiltonian, the left or right eigenstates alone do not satisfy the orthogonality condition; both the left and right
	eigenstates should be used which satisfy the bi-orthogonal relation \cite{Brody_2014}, i.e.,
	$\langle\psi_n^{L}|\psi_n^{R}\rangle=\delta_{mn}$,
	where $H|\psi_n^{R}\rangle=E_n|\psi_n^{R}\rangle$, $H^{\dag}|\psi_n^{L}\rangle=E_n^{\ast}|\psi_n^{L}
	\rangle$, $n=1,2,\dots,N$. Incorporating it into Eq. \eqref{eqs23}, the LDOS at the $l$-th site of the 1D vertical chain model can be expressed as
	\begin{equation}\label{eqs25}
		\begin{split}
			\rho_l(\epsilon_r,\epsilon_i)=&\frac{1}{N}\sum_n\delta(\epsilon_r-\mathrm{Re}[E_n])\delta(\epsilon_i-\mathrm{Im}[E_n])\\
			&\times|\langle l|\psi^R_n\rangle\langle\psi^L_n|l\rangle|,
		\end{split}
	\end{equation}
	where $\rho_l(\epsilon_r,\epsilon_i)$ is defined on the complex energy plane, where $\epsilon_r$ ($\epsilon_i$) represents the real (imaginary) part of the complex energy $\epsilon$. In our numerical calculations, since the 1D vertical chain model has sublattice degrees of freedom, $N=2L_y$ with $L_y$ is the 1D chain length, and the phase factor $\langle l |\psi_n^{R}\rangle \langle\psi_n^{L}| l \rangle$ includes the sum of the $(2l-1,2l-1)$ and $(2l,2l)$ matrix elements of the $2L_y\times2L_y$ matrix. Besides, in the plot of the LDOS, the $\delta$-peaks are broadened by a finite width controlled by a factor $b$. This is achieved by treating the Dirac-$\delta$ function as a Lorentzian form $\delta(\omega-\omega_n)\to \frac{1}{2\pi}\frac{b}{(\omega-\omega_n)^2+b^2}$.
	
	\vspace{2ex}
	\noindent
	{\bf Time evolution of Gaussian wave-packet.} We set the initial state as a 2D Gaussian wave-packet
	\begin{equation}\label{eqs28} \psi(t=0)=\frac{1}{(4\pi\sigma^2)^{\frac{1}{2}}}\mathrm{exp}[-\frac{(x-x_0)^2}{4\sigma^2}-\frac{(y-y_0)^2}{4\sigma^2}](1,1)^T,
	\end{equation}
	where $(x_0,y_0)$ represents the center of wave-packet. According to the Schrödinger equation $i\partial_t|\psi(t)\rangle=H|\psi(t)\rangle$, the
	time-evolution operator of the non-Hermitian system is expressed as
	\begin{equation}\label{eqs27}
		U(t)=e^{-iHt}=\sum_n e^{-iE_nt}|\psi^R_n\rangle\langle\psi^L_n|,
	\end{equation}
	which shows that those modes with $\mathrm{Im}[E_n]<0$ will vanish due to the exponentially decaying factor, whereas those with $\mathrm{Im}[E_n]>0$ will dominate in the long-time limit. For the model of Chern insulator with a non-Hermitian boundary in our work, since the term $H_{\gamma}$ contains an imaginary unit factor, the total Hamiltonian satisfies $H^*(\gamma)=H_0-H_{\gamma}=H_0+
	H_{-\gamma}=H(-\gamma)$. So the sign of $\gamma$ determines the sign of $\mathrm{Im}[E_n]$, which in turn determines whether the mode amplitude grows or decays.
	
	\vspace{2ex}
	\noindent
	{\bf Realization of the non-Hermitian boundaries via coupling to environments.}
	In open quantum systems, the coupling between the system and the environment can be described by Lindblad master equation, i.e.,
	\begin{equation}\label{eqs32}
		\frac{\mathrm{d}\rho}{\mathrm{d}t}=-i[H,\rho]+\sum_{\mu}(2L_\mu\rho L_\mu^\dag-\left\{L_\mu^\dag L_\mu,\rho\right\}),
	\end{equation}
	where $\rho$ is the density matrix, $L_\mu$'s are the Lindblad dissipators describing quantum jumps due to coupling to the environment. The short-time evolution is described by the Schr\"{o}dinger evolution under the effective non-Hermitian Hamiltonian $H_{eff}=H-i\sum_{\mu}{L_\mu^\dag L_\mu}$ with $\mathrm{d}\rho/\mathrm{d}t=-i(H_{eff}\rho-\rho H_{eff}^\dag)$ \cite{song2019non1,nakagawa2018non}. Considering the single particle loss and gain with the loss and gain dissipators:
	\begin{equation}\label{eqs33}
		\begin{aligned}
			L_\mu^l&=\sqrt{\gamma_l}(c_{\mu a}+c_{\mu b}),\\
			L_\mu^g&=\sqrt{\gamma_g}(c_{\mu a}^\dag+c_{\mu b}^\dag).
		\end{aligned}
	\end{equation}
	Eq. \eqref{eqs33} leads to the effective Hamiltonian, which reads in momentum space as
	\begin{equation}\label{eqs34}
		H_{eff}=H+i\left(\gamma_g-\gamma_l\right)\left(\sigma_0+\sigma_x\right).
	\end{equation}
	The second term in Eq. (\ref{eqs34}) has a form that produces the non-Hermitian boundary studied in Eq. \eqref{eq2} and Eq. \eqref{eqs29}. The loss and gain dissipators in Eq. \eqref{eqs33} can be realized in reservoir engineering  by using nonlocal couplings to auxiliary degrees of freedom which undergo local loss or gain \cite{gong2018topological}.
	
	\vspace{2ex}
	\noindent
	\textbf{Data availability}
	
	\noindent
	The data generated and analyzed during this study are available from the corresponding author upon reasonable request.
	
	\vspace{2ex}
	\noindent
	\textbf{Code availability}
	
	\noindent
	All code used to generate the plots within this paper are available from the corresponding author upon reasonable request.
	
	\vspace{2ex}
	\noindent
	\textbf{Acknowledgments}
	
	\noindent
	This work was supported by National Natural Science Foundation of China (Grant No. 12274206,  No.12034014 and No.11904245),  the Innovation Program for Quantum Science and Technology (Grant No. 2021ZD0302800), National Key  R\&D Program of China (Grant No. 2022YFA1403601), and the Xiaomi foundation.
	
	\vspace{2ex}
	\noindent
	\textbf{Author contributions}
	
	\noindent
	All authors performed the calculations, discussed the results and prepared the manuscript.
	
	\vspace{2ex}
	\noindent
	\textbf{Competing interests}
	
	\noindent
	The authors declare no competing financial interests.

\end{document}